\documentclass[preprint]{revtex4-1}
\usepackage{graphicx}
\usepackage{dcolumn}
\usepackage{setspace}
\usepackage{hyperref}
\usepackage{natbib}
\usepackage[utf8]{inputenc}

\begin{document}
\title{Anisotropic magnetic properties of URhIn$_{5}$ compound}

\author{\underline{Attila Bartha}}
 \email[Corresponding author: ]{bartha@mag.mff.cuni.cz}
 \affiliation{Department of Condensed Matter Physics, Charles University, Ke Karlovu 5, 121 16 Praha 2, Czech Republic}
\author{M. Kratochv\'{i}lov\'{a}}  
 \affiliation{Department of Condensed Matter Physics, Charles University, Ke Karlovu 5, 121 16 Praha 2, Czech Republic}
\author{V. Sechovsk\'{y}}
 \affiliation{Department of Condensed Matter Physics, Charles University, Ke Karlovu 5, 121 16 Praha 2, Czech Republic}
\author{J. Custers}
 \affiliation{Department of Condensed Matter Physics, Charles University, Ke Karlovu 5, 121 16 Praha 2, Czech Republic}
\author{M. Du\v sek}
 \affiliation{Department of Structure Analysis, Institute of Physics ASCR, Cukrovarnick\'{a} 10, 162 00 Praha 6, Czech Republic} 
 
\begin{abstract}
We report on synthesis and anisotropic physical properties of URhIn$_{5}$. High quality single crystals were grown in In-flux. The compound undergoes a second order phase transition into an antiferromagnetic state at \textit{T}$_{\textrm{\tiny{N}}}$ = 98 K. The transition is field independent up to 9 T. An increase of the resistivity $\rho$ with \textit{j} along the [100], [110] and [001] tetragonal axis indicates a spin-density-wave induced order with the gap opening first along the [001] direction. The magnetic susceptibility \mbox{$\chi$ = \textit{M}/\textit{H}} exhibits a strong anisotropy. Above \textit{T} $=$ 200 K, $\chi$(\textit{T}) follows Curie-Weiss law with the effective moment of $\mu_{\textrm{\tiny{eff}}}$ = 3.71 $\mu_{\textrm{\tiny{B}}}/$U and the Weiss temperatures of $\theta_{\textrm{\tiny{P}}}^{[100]} = -900$ K and \mbox{$\theta_{\textrm{\tiny{P}}}^{[001]} = -500$ K} for \textit{H} $\parallel$ [100] and \textit{H} $\parallel$ [001] respectively. The characteristic Kondo-like temperature for URhIn$_{5}$ yields \textit{T}$_{\textrm{\tiny{K}}} =$ 125 K. 
\end{abstract}

\maketitle

\begin{description}
\item[PACS number(s)]{75.30.Gw, 75.50.Ee, 81.10.Dn}
\end{description}

\section{Introduction}
In an early manuscript of Monthoux \textit{et al.} \cite{monthoux} it was noted that a tetragonal lattice structure is a preferred host for unconventional superconductivity where pairing of the Cooper-pairs is mediated by antiferromagnetic spin fluctuations. It was further pointed out that increasing dimensionality from 2D (tetragonal) to 3D (cubic structure) while keeping otherwise same conditions, would lead to a decrease of \textit{T}$_{\textrm{\tiny{C}}}$. Exemplary for their magnetic interaction model is the Ce$_{n}$\textit{T}In$_{3n+2}$ group of compounds where \textit{T} stands for a transition element (Co, Rh, Ir) and \textit{n} signifies the dimension of the CeIn$_{3}$ structure. \textit{n} = $\infty$ denotes the cubic (3D) CeIn$_{3}$. With decreasing \textit{n} the dimensionality becomes more 2D. For instance, CeCoIn$_{5}$ shows a superconducting transition temperature exceeding 2 K. Inserting an additional layer of CeIn$_{3}$ into CeCoIn$_{5}$ gives Ce$_{2}$CoIn$_{8}$. This compound is more 3D. In line with the model \textit{T}$_{\textrm{\tiny{C}}}$ is reduced being only \mbox{0.4 K} \cite{genfu}. Furthermore, the pure 3D compound CeIn$_{3}$ shows superconductivity at even lower temperature, i.e, below \textit{T}$_{\textrm{\tiny{C}}}$ = 0.2 K \cite{cein3}.
Recently the Ce$_{n}$\textit{T}In$_{3n+2}$ compounds come into focus for being excellent candidates for studying the effect of dimensionality on the ground state properties. The parameter \textit{dimensionality} plays a key role in the global phase diagram of heavy fermions \cite{jonas}.

While the cerium 4\textit{f} electron in general is strongly bound to the core, typically the properties of the actinide compounds are characterized by the large spatial extent of the 5\textit{f} wave function. A comparison with Ce counterpart can shed light on the influence of strong hybridization of U 5\textit{f}-electron states of valence electrons of ligands. 

Among the ternary uranium compounds URhIn$_{5}$ is the first known representative of this group. The material orders antiferromagnetically below \textit{T}$_{\textrm{\tiny{N}}}$ = 98 K. The Sommerfeld coefficient is weakly enhanced yielding $\gamma = 50$ mJ$\cdot$mol$^{-1}\cdot$K$^{-2}$ \cite{matsumoto}. The ground state properties of URhIn$_{5}$ resemble those of UIn$_{3}$ parent compound \cite{uin3}. In this paper we report on the single crystal growth and physical properties of this novel material, we focus on the anisotropy of the magnetic properties in detail.

\section{Experimental}
Single crystals of URhIn$_{5}$ were grown using In self-flux method. High quality elements U (purified by SSE), Rh (3N5) and In (5N) with starting composition of \mbox{U:Rh:In = 1:1:25} were placed in an alumina crucible. The crucible was further sealed in an evacuated quartz tube. The ampoules were then heated up to 1050 $^\circ$C, kept at this temperature for 10 h to homogenize the mixture properly and consequently cooled down to 750 $^\circ$C in 100 h. After decanting, plate-like single crystals with typical dimensions of \mbox{$1\times 1\times 0.5$ \textrm{mm}$^{3}$} were obtained. 

Homogeneity and chemical composition of the single crystals were confirmed by scanning electron microscope (Tescan MIRA I LMH SEM). The apparatus is  equipped with energy dispersive \textit{X}-ray analyzer (Bruker AXS). The crystal structure was determined by \textit{X}-ray powder diffraction (Bruker D8 Advance diffractometer) and single crystal \textit{X}-ray diffraction (Rigaku Rapid). The obtained diffraction patterns were refined according to standard Rietveld technique using FullProf/WinPlotr software \cite{fullprof}. The \textit{X}-ray analysis revealed the HoCoGa$_{5}$-type structure with lattice parameters \mbox{\textit{a} = 4.6159(2) \AA} and \mbox{\textit{c} = 7.4120(6) \AA} which corresponds to previously published values by Matsumoto and co-workers \cite{matsumoto}. 

The electrical resistivity measurements were done utilizing standard four-point method down to 2 K in a Physical Property Measurement System (PPMS). The specific heat measurements were carried out using the He3 option of the PPMS. Temperatures as low as 400 mK were reached. Magnetization measurements were performed in a superconducting quantum interference device from 2 to 300 K and magnetic fields \mbox{up to 7 T.}

\section{Results and discussion}
\begin{figure}[h!]
 \includegraphics[width=0.475\textwidth, keepaspectratio=true]{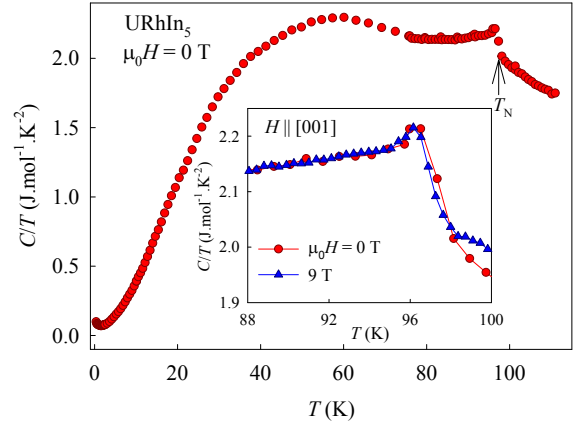}
 \caption{\label{fig:specific heat}Temperature dependence of the specific heat divided by temperature. The transition into the antiferromagnetic state at \textit{T}$_{\textrm{\tiny{N}}}$ = 98 K is marked by an arrow. Inset: comparison of \textit{C}/\textit{T} in zero and in applied field of 9 T along the [001] axis.}
 \end{figure}

FIG. \ref{fig:specific heat} shows the temperature dependence of the specific heat \textit{C}(\textit{T}) divided by temperature. The clear $\lambda$-shaped anomaly at \textit{T}$_{\textrm{\tiny{N}}}$ = 98 K indicates a second order phase transition in agreement with Ref. \cite{matsumoto}. However, contrary to those data, no additional feature in \textit{C}(\textit{T}) at \textit{T} $\sim$ 40 K is observed attesting measurement on single phase sample. The magnitude for the phonon contribution was determined from a $C/T = \gamma + \beta T^2$ fit to the data (fit interval 1 K $<$ \textit{T} $<$ 10 K). The value of Sommerfeld coefficient yields $\gamma$ = 60.7 mJ$\cdot$mol$^{-1}\cdot$U$\cdot$K$^{-2}$. The value of $\beta$ coefficient equals to 3.3 mJ$\cdot$mol$^{-1}\cdot$U$\cdot$K$^{-4}$ and corresponds to a Debye temperature \textit{T}$_{\textrm{\tiny{D}}}$ = 165 K. These values are close to those presented in Ref. \cite{matsumoto}. The inset of FIG. \ref{fig:specific heat} presents data in applied magnetic field of 9 T along \textit{c}-axis. The direct comparison with zero field measurement reveals that the position of \textit{T}$_{\textrm{\tiny{N}}}$ is almost unaffected within experimental uncertainty.
Interestingly, the CeRhIn$_{5}$ counterpart is rather insensitive to the application of magnetic field along the same direction as well \cite{cerhin5}.

\begin{figure}
 \includegraphics[width=0.475\textwidth, keepaspectratio=true]{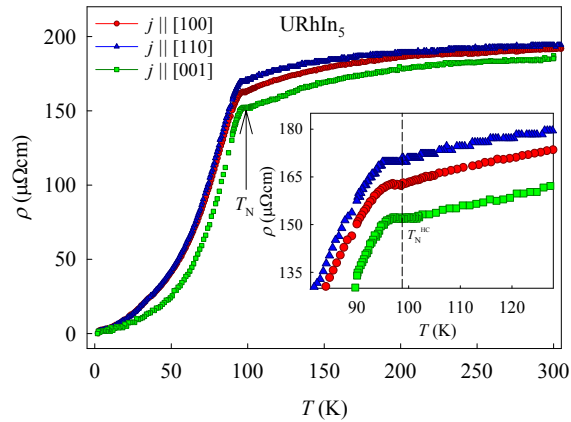}
 \caption{Temperature dependence of the electrical resistivity for current applied along [100], [110] and [001] directions. The arrows mark the onset of the anomalies. The inset shows the transition in more detail. Dashed line shows \textit{T}$_{\textrm{\tiny{N}}}$ obtained from heat capacity measurements.}
 \label{fig:resistivity}
\end{figure}

FIG. \ref{fig:resistivity} summarizes the overall temperature dependence of the electrical resistivity for current \textit{j} applied along the [100], [110] and [001] axes. The room temperature resistivity equals 180 $\mu\Omega$cm in the basal plane and is only slightly lower for the \textit{c}-axis direction \mbox{$\rho^{[001]}$ = 170 $\mu\Omega$cm}. Note that $\rho^{[100]}$ (300 K) is almost the same as reported by Matsumoto and co-workers \cite{matsumoto}. Above \textit{T}$_{\textrm{\tiny{N}}}$, the resistivity in all directions shows rather weak temperature dependence with positive "metallic" temperature coefficient d$\rho$/d\textit{T}. The resistivity exhibits a smooth change from high temperature concave-like decrease to low-temperature convex one with inflection point at \mbox{$\sim$ 125 K}.
The data manifest distinct anomalies with onset at around $\sim$ 100 K for $\rho^{[001]}$ and at a slightly lower temperature \textit{T} = 98 K for $\rho^{[100]}$ and $\rho^{[110]}$, respectively, which is reminiscent of the N\'{e}el temperature anomaly for $\rho$(\textit{T}) in pure Cr \cite{chromium} - a spin-density-wave (SDW) antiferromagnet. Accordingly, the onset marks \textit{T}$_{\textrm{\tiny{N}}}$ and the increase in resistivity results from opening of the SDW gap. Hence, the higher onset temperature of the anomaly in $\rho^{[001]}$ strongly suggests that opening of the gap occurs primarily along this direction. Below \textit{T}$_{\textrm{\tiny{N}}}$ the resistivity in all directions drops rapidly. The temperature dependence of $\rho$ down to lowest \textit{T} can be fitted using the equation appropriate for an energy gap ($\Delta$) antiferromagnet with and additional \textit{T}$^{2}$ Fermi liquid term, being: $\rho(T)=\rho_{0}+AT^{2}+DT(1+2T/\Delta)\textrm{exp}(-\Delta/T)$ \cite{resistivity}.
Best fitting of $\rho^{[100]}$ gives a residual resistivity value \mbox{$\rho_{0}$ = 1 $\mu\Omega$cm}, an electron-electron scattering coefficient of \textit{A} = 0.013 $\mu \Omega$cm$\cdot$K$^{-2}$, an electron-magnon and spin-disorder scattering prefactor \textit{D} = 0.35 $\mu \Omega$cm$\cdot$K$^{-1}$ and \mbox{$\Delta$ = 82 K}. Our fit yielded similar values of the parameters for current along [110] direction. However in the case of \mbox{\textit{j} $\parallel$ [001]} we obtained a somewhat higher value of \mbox{$\Delta$ = 119 K}, which supports our speculation of a gap opening at higher temperatures.

\begin{figure}
 \includegraphics[width=0.475\textwidth, keepaspectratio=true]{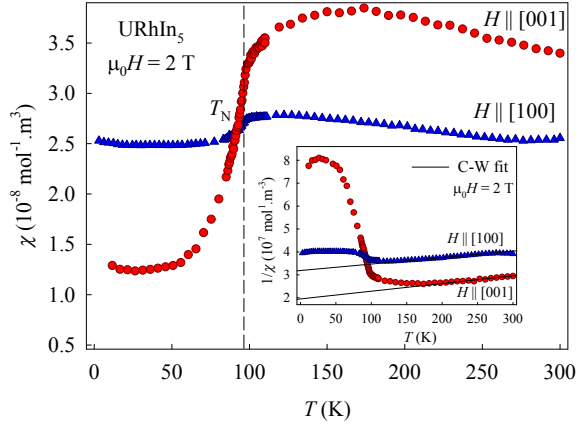}
 \caption{Temperature dependence of magnetic susceptibility for a magnetic field oriented along the [100] and [001] direction. Vertical dashed line marks the transition temperature \textit{T}$_{\textrm{\tiny{N}}}$ obtained from specific heat. The inset shows the inverse susceptibility and Curie-Weiss fit for magnetic field oriented along [001] and [100] directions.}
 \label{fig:magnetization}
\end{figure}

FIG. \ref{fig:magnetization} shows the temperature dependence of the magnetic susceptibility in a magnetic field of 2 T oriented along [100] and [001]. Our data resemble the susceptibility data presented in Ref. \cite{matsumoto}. The inverse magnetic susceptibility follows the Curie-Weiss law above 200 K with an effective magnetic moment \mbox{$\mu_\textrm{\tiny{eff}}$ = 3.71 $\mu_{\textrm{\tiny{B}}}$/U}, which is very close to the U$^{3+}$ ion value (3.62 $\mu_{\textrm{\tiny{B}}}$). Rather more striking are the large negative Curie-Weiss temperatures $\theta_{\textrm{\tiny{P}}}^{[001]}$ = $-500$ K for \mbox{\textit{H} $\parallel$ \textit{c}} and $\theta_{\textrm{\tiny{P}}}^{[100]}$ = $-900$ K for \mbox{\textit{H} $\parallel$ \textit{a}}, indicative strong predominant antiferromagnetic spin exchange. The significant difference between the values $\theta_{\textrm{\tiny{P}}}^{[001]}$ and $\theta_{\textrm{\tiny{P}}}^{[100]}$ points to strong magnetocrystalline anisotropy in URhIn$_{5}$. The observed shallow maximum in $\chi$(\textit{T}) for \textit{H} $\parallel$ [001] at \mbox{$\sim$ 150 K} is in agreement with that observed before \cite{matsumoto} and discussed in context of a characteristic Kondo scale. When assuming that Kondo physics applies in URhIn$_{5}$ and neglecting crystal field influence the Kondo temperature amounts to \mbox{\textit{T}$_{\textrm{\tiny{K}}}$ = $\vert\theta_\textrm{\tiny{P}}\vert/4$ giving 125 K} when inserting $\theta_{\textrm{\tiny{P}}}^{[001]}$ \cite{hewson}. This temperature coincides the inflection point in $\rho$(\textit{T})$^{[001]}$ suggesting a common origin of both.
Noteworthy, the $\chi$(\textit{T}) curve for \textit{H} $\parallel$ [100] at low temperatures \textit{T} $<$ 40 K differs significantly from published data \cite{matsumoto}. The increase of the susceptibility, as is seen in FIG. \ref{fig:magnetization} is much less pronounced. This observation could point to an extrinsic nature of the upturn which means a smaller amount of impurities in our sample compared to Ref. \cite{matsumoto}.

\section{Conclusions}
High quality single crystals of the antiferromagnetic URhIn$_{5}$ compound were successfully synthesized. Our results of specific heat and magnetic susceptibility measurements are in good agreement with previous results of Matsumoto and co-workers \cite{matsumoto}. The compound orders antiferromagnetically below \textit{T}$_{\textrm{\tiny{N}}}$ = 98 K. Resistivity data reveal anisotropy in the spin-gap structure, with larger gap along [001] crystal axis. In addition, we showed that \textit{T}$_{\textrm{\tiny{N}}}$ is robust against magnetic field up to 9 T. Furthermore, the compound exhibits a strong magnetocrystalline anisotropy. URhIn$_{5}$ is the first representative of the layered tetragonal family of materials based on UIn$_{3}$ building blocks. Recalling the plenitude of intriguing phenomena reported in Ce$_{n}$\textit{T}In$_{3n+2}$ compounds the investigation of hypothetical U$_{2}$\textit{T}In$_{8}$ and U\textit{T}In$_{5}$ materials is highly desirable. 

\section{Acknowledgement}
This work was supported by the Grant Agency of the Charles University (project no. 362214). Experiments were performed in MLTL (http://mltl.eu/), which is supported within the program of Czech Research Infrastructures (project no. LM2011025).

\end{document}